\PassOptionsToPackage{dvipsnames}{xcolor}

\documentclass[sigconf,9pt]{acmart}
\AtBeginDocument{%
  \providecommand\BibTeX{{%
    \normalfont B\kern-0.5em{\scshape i\kern-0.25em b}\kern-0.8em\TeX}}}

\acmYear{2021}\copyrightyear{2021}
\setcopyright{acmcopyright}
\acmConference[SIGCOMM '21]{ACM SIGCOMM 2021 Conference}{August 23--28, 2021}{Virtual Event, USA}
\acmBooktitle{ACM SIGCOMM 2021 Conference (SIGCOMM '21), August 23--28, 2021, Virtual Event, USA}
\acmPrice{15.00}
\acmDOI{10.1145/3452296.3472933}
\acmISBN{978-1-4503-8383-7/21/08}

\usepackage[utf8]{inputenc}
\usepackage{graphicx}
\usepackage{subcaption}
\usepackage{fancyvrb}
\usepackage{multirow}
\usepackage{makecell}
\usepackage{array, multirow}
\usepackage{arydshln}
\usepackage{balance}
\usepackage[ruled,vlined]{algorithm2e}

\PassOptionsToPackage{hyphens}{url}
\PassOptionsToPackage{obeyspaces}{url}
\PassOptionsToPackage{spaces}{url}

\usepackage{enumitem}
\setlist{noitemsep}

\usepackage{pifont}
\newcommand{\cmark}{\ding{51}} %
\newcommand{\xmark}{\ding{55}} %

\newcolumntype{P}[1]{>{\centering\arraybackslash}p{#1}}

\newcommand{\ignore}[1]{}
\newcommand{\sad}{{\sf SadDNS}}
\newcommand{\frag}{{\sf FragDNS}}
\newcommand{\hijack}{{\sf HijackDNS}}
\newcommand{\saddns}{{\sf SadDNS}}

\newenvironment{added}{ }{ }

\usepackage{array}
\newcolumntype{H}{@{}>{\setbox0=\hbox\bgroup}c<{\egroup}}

\newcommand{\mr}[2]{ \multirow{#1}{*}{#2} }             %
\newcommand{\mcc}[2]{ \multicolumn{#1}{c|}{#2} }        %

\usepackage{tikz}

\usepackage[printwatermark]{xwatermark}
\newwatermark[allpages,color=black!100,angle=0,scale=1,xpos=0,ypos=-126]{\small SIGCOMM '21: Proceedings of the 2021 ACM SIGCOMM 2021 Conference, August 2021, Pages 836-849 \\
Accepted version. \url{https://doi.org/10.1145/3452296.3472933}}

\begin{document}

\title{From IP to Transport and Beyond:\\Cross-Layer Attacks Against Applications}

\settopmatter{authorsperrow=4}
\author{Tianxiang Dai}
\affiliation{%
  \institution{Fraunhofer SIT}
  \country{Germany}
}
\author{Philipp Jeitner}
\affiliation{%
  \institution{Fraunhofer SIT}
  \institution{TU Darmstadt}
  \country{Germany}
}
\author{Haya Shulman}
\affiliation{%
  \institution{Fraunhofer SIT}
  \country{Germany}
}
\author{Michael Waidner}
\affiliation{%
  \institution{Fraunhofer SIT}
  \institution{TU Darmstadt}
  \country{Germany}
}



\begin{abstract}
We perform the first analysis of methodologies for launching DNS cache poisoning: manipulation at the IP layer, hijack of the inter-domain routing and probing open ports via side channels. We evaluate these methodologies against DNS resolvers in the Internet and compare them with respect to effectiveness, applicability and stealth. Our study shows that DNS cache poisoning is a practical and pervasive threat. 

We then demonstrate cross-layer attacks that leverage DNS cache poisoning for attacking popular systems, ranging from security mechanisms, such as RPKI, to applications, such as VoIP. In addition to more traditional adversarial goals, most notably impersonation and Denial of Service, we show for the first time that DNS cache poisoning can even enable adversaries to bypass cryptographic defences: we demonstrate how DNS cache poisoning can facilitate BGP prefix hijacking of networks protected with RPKI even when all the other networks apply route origin validation to filter invalid BGP announcements. Our study shows that DNS plays a much more central role in the Internet security than previously assumed.

We recommend mitigations for securing the applications and for preventing cache poisoning.
\end{abstract}

\maketitle

\section{Introduction}
Domain Name System (DNS), [RFC1034, RFC1035] \cite{rfc1034,rfc1035}, plays a central role in the Internet. Designed and standardised in the 80s to provide lookup services DNS has evolved into a complex infrastructure and is being increasingly used to support a wide variety of existing and future applications and security mechanisms. Given the large dependency of the Internet on DNS it also became a lucrative target for attacks.

{\bf DNS cache poisoning.} In a cache poisoning attack an adversary injects malicious DNS records into the cache of a victim DNS resolver. Poisoning the cache enables the adversary to redirect the victims using that DNS resolver to malicious hosts instead of the genuine servers of the target domain. As a result, the adversary intercepts all the services in the target domain.

In this work we explore how practical off-path DNS cache poisoning attacks are and how such attacks can be exploited to launch cross-layer attacks against applications.

{\bf Taxonomy of cache poisoning methodologies.} As we explain in Section \ref{app:sc:bck}, off-path DNS cache poisoning is challenging to launch in practice. Nevertheless, there are methodologies that, depending on different conditions, can result in practical attacks. In this work we evaluate such methodologies for launching cache poisoning attacks:
(1) BGP prefix hijacking, (2) transport layer side channels and (3) injections into IP defragmentation cache. These methodologies were previously used for issuing fraudulent certificates, \cite{birge2018bamboozling} or for hijacking bitcoins \cite{apostolaki2017hijacking}. Attacks for issuing fraudulent certificates were also carried out by \cite{brandt2018domain} using IP fragmentation; a method initially proposed in \cite{cns:frag:dns}. \cite{saddns} combined ICMP error messages and rate limiting of nameservers to create a side channel for guessing the source port in DNS requests, but have not evaluated this attack against real Internet systems.

{\em Which of the methods is more effective? Which has higher applicability? Which is stealthier and does not trigger alerts?}

To answer these questions we perform the first comparative analysis of the methodologies for cache poisoning attacks. In addition, in order to gain a deeper understanding of the methodologies and their impact on the Internet applications, we also extend the evaluations in the previous work \cite{birge2018bamboozling,brandt2018domain,saddns} for Internet scale measurements of applicability and effectiveness of these methodologies against multiple Internet networks.

{\bf Taxonomy of vulnerable applications.} The implications of cache poisoning for the other Internet services and applications has not been explored. There is evidence of cache poisoning in the wild, mostly for redirecting victims to impersonating websites \cite{myetherwallet}. Cache poisoning was also demonstrated in research against the certificate authorities \cite{birge2018bamboozling,brandt2018domain}. But there is no comprehensive study of exploits of cache poisoning against Internet clients and services.

{\em What applications are at risk due to cache poisoning? How can an attacker exploit cache poisoning to attack applications? What is the fraction of vulnerable applications in the Internet? What are the challenges and what cache poisoning methodologies are more suitable?}

We answer these questions by evaluating the cache poisoning methodologies against a range of popular applications. We defined nine categories of applications, ranging from security mechanisms, to VoIP, email and intermediate devices; see Table \ref{tab:dns_systems}. We provide the first systematic study of cache poisoning against a collection of popular applications and security mechanisms.

{\bf Poisoning is a threat to applications.} Our results demonstrate that, {\em although challenging to launch, off-path DNS cache poisoning poses a realistic threat for many Internet applications}. Surprisingly, we show that DNS cache poisoning can be applied for downgrade attacks against security mechanisms causing the victims not to perform validation, e.g., RPKI or domain-based anti-spam validation. Taking RPKI as an example, we developed an attack that by redirecting the RPKI cache [RFC6810] \cite{rfc6810} to a wrong repository via DNS cache poisoning, the attacker can cause the RPKI validation to result in status {\tt unknown} (instead of {\tt invalid}). As a result the RPKI cache will not validate correctness of the BGP announcements that it receives. Suppressing RPKI validation allows the adversary to perform BGP prefix hijacks even of ASes which have the corresponding RPKI material (Route Origin Authorization and resource certificates \cite{rfc6482}) in the public repositories and hijack even the senders which enforce route origin validation \cite{rfc6811}. 

Another example is malware distribution by causing the anti-spam validation to fail via cache poisoning.

This is the first demonstration of the devastating power of DNS cache poisoning, which shows that in addition to traditional threats, such as impersonation, DNS cache poisoning can facilitate much stronger attacks which were otherwise not possible. We also show that DNS cache poisoning can be used to inflict Denial of Service (DoS) on applications and their clients.

In our experimental evaluation against the applications we exploit DNS cache poisoning to subvert correctness and security of basic Internet functions, enabling the attackers to take over IP addresses, to hijack telephony, to de-synchronise local time, and even prevent victims from connecting to the correct VPN tunnel. %

{\bf Off-path attacks.} Our study is performed with off-path attackers. This is the weakest attacker model in the Internet, it can merely send packets from spoofed IP addresses, which is a realistic assumption since around 30\% of the Internet networks do not enforce egress filtering \cite{beverly2005spoofer, beverly2009understanding, mauch2013open,beverly2013initial,lone2018using,luckie2019network}. Essentially any adversary in the Internet has off-path capabilities and can select networks which allow it to send packets with spoofed source IP addresses. Stronger attackers, most notably the on-path Man-in-the-Middle (MitM), can do more devastating attacks. Nevertheless, MitM attackers are more rare and even such attackers have limitations: the strong government sponsored attackers can be on-path only to some of the Internet victims depending on the paths that they control but even they do not control all of the networks. Therefore, it is critical to understand the threat that an off-path attacker poses to applications.

{\bf Disclosure and ethics.} Our attacks were tested against remote networks reliably, yet were ethically compliant. We measured and evaluated vulnerabilities in the DNS caches of the subjects of our study and measured which services use the caches but did not hijack their traffic nor Internet resources and neither did we place incorrect DNS records for Internet domains that are not under our control in the caches of our test subjects. Specifically, to avoid harming Internet customers and domains, we set up a victim AS and victim domains as well as adversarial AS and adversarial hosts on that AS, which were used by us for carrying out the attacks against the victims. Our measurement study for evaluating the vulnerabilities was performed using our victim domains, which ensured that the targets of our study would not use the spoofed records for any ``real'' purpose. %

We believe that in addition to disclosing the vulnerabilities to the affected entities it is critical to raise awareness to the extent and the scope of the vulnerabilities. 

{\bf Contributions.} We present the first comprehensive study of the attack surface that off-path DNS cache poisoning introduces on the Internet ecosystem. 

$\bullet$ We implement three methodologies for launching off-path DNS cache poisoning attacks: (1) BGP prefix hijacking, (2) side-channels and (3) fragmentation. We perform the first Internet-scale evaluation of these methodologies against DNS resolvers and compare them for applicability, stealthiness and success of cache poisoning.

$\bullet$ We apply these methodologies to launch {\em cross-layer attacks} against widely used applications and services (see taxonomy in Table \ref{tab:dns_systems}). Our study shows that cache poisoning can be used to bypass security mechanisms, to cause DoS attacks, or for impersonation attacks.

$\bullet$ We provide recommendations for countermeasures for DNS caches against cache poisoning attacks and for applications against cross-layer attacks even when using poisoned caches.%

{\bf Organisation.} We review DNS cache poisoning and related work in Section \ref{app:sc:bck}. In Section \ref{sc:dns:poison} we present the DNS cache poisoning methodologies that we use throughout our work. In Section \ref{sc.taxonomy} we demonstrate cross-layer attacks against applications using DNS cache poisoning. We provide results of our measurements in Section \ref{sc:eval} and recommend mitigations in Section \ref{sc:defences}. We conclude this work in Section \ref{sc:conc}. 

\section{DNS Cache Poisoning Overview}
\label{app:sc:bck}

Domain Name System (DNS) \cite{rfc1035} cache poisoning allows an attacker to redirect victims to attacker controlled hosts. Typically the attackers targets recursive DNS resolvers whose caches serve multiple clients. A single injection of a malicious DNS record propagates to all the hosts that use that resolver. The attacker can then intercept the traffic between the services (such as web, email, FTP) in the victim domain and the hosts that use the poisoned cache. DNS resolvers use defences to make launching successful cache poisoning attacks difficult.

\subsection{Defences Against Poisoning} The DNS resolvers are required to randomise certain fields in DNS requests sent to the nameservers, [RFC5452] \cite{rfc5452}. These include a random 16 bit UDP source port and the 16 bit DNS transaction identifier (TXID); additional defences include nameserver randomisation \cite{rfc5452} and 0x20 encoding \cite{DagonAVJL08}. The nameservers copy these fields from the DNS request to the DNS response. DNS resolvers accept the first DNS response with the correctly echoed challenge values and ignore any responses with incorrect values. %

To launch a successful cache poisoning attack, the attacker needs to guess the correct challenge values and make sure that his spoofed response arrives before the genuine response from the real nameserver. %
This is easy for an on-path (man-in-the-middle) attacker, which can simply copy the values from the request to the response. Cryptographic signatures with DNSSEC [RFC6840] \cite{weiler2013rfc} could prevent on-path attacks, however, DNSSEC is not widely deployed. Less than 1\% of the second level domains (e.g., 1M-top Alexa) are signed, and most resolvers do not validate DNSSEC signatures, e.g., \cite{chung2017longitudinal} found only 12\% in 2017. Our measurements indicate that less than 5\% of the domains we studied are signed. There is however an increase in the resolvers validating DNSSEC: we found 28.6\% validating resolvers via our ad-network study. Deploying DNSSEC was shown to be cumbersome and error-prone \cite{chung2017understanding}. Even when widely deployed DNSSEC may not always provide security: a few research projects identified vulnerabilities and misconfigurations in DNSSEC deployments in popular registrars \cite{shulman2017one,jeitner2021injection}. 

Recent proposals for encryption of DNS traffic, such as DNS over HTTPS \cite{hoffman2018rfc} and DNS over TLS \cite{hu2016rfc}, although vulnerable to traffic analysis \cite{shulman2014pretty,siby2019encrypted}, may also enhance resilience to cache poisoning. These mechanisms are not yet in use by the nameservers in the domains that we tested. Nevertheless, even if they become adopted, they were not designed to protect the entire resolution path, but only the link between the client and the recursive resolver, and hence will not prevent DNS cache poisoning attacks.

\subsection{History of DNS Cache Poisoning}
In 2007 Klein identified vulnerability in Bind9 DNS resolvers \cite{klein2007bind} and in Windows DNS resolvers \cite{klein2007windows} allowing off-path attackers to reduce the entropy introduced by the TXID randomisation. In 2008 Kaminsky \cite{Kaminsky08} presented a practical cache poisoning attack even against truly randomised TXID. Vixie suggested to randomise the UDP source ports already in 1995 \cite{Vixie95}, subsequently in 2002 Bernstein warned that relying on randomising TXID alone is vulnerable \cite{Bernstein:DNS}. Following Kaminsky attack DNS resolvers were patched against cache poisoning \cite{rfc5452}, and most randomised the UDP source ports in queries. %

Nevertheless, shortly after new approaches were developed allowing cache poisoning attacks. In 2012 \cite{conf/esorics/HerzbergS12} showed that off-path attackers can use side-channels to infer the source ports in DNS requests. In 2015 \cite{shulman2015towards} showed how to attack resolvers behind upstream forwarders. This work was subsequently extended by \cite{zheng2020poison} with poisoning the forwarding devices. A followup work demonstrated such cache poisoning attacks also against stub resolvers \cite{collaborative-poison}. \cite{saddns} showed how to use ICMP errors to infer the UDP source ports selected by DNS resolvers. Recently \cite{klein2020cross} showed how to use side channels to predict the ports due to vulnerable PRNG in Linux kernel. In 2013 \cite{cns:frag:dns} provided the first feasibility result for launching cache poisoning by exploiting IPv4 fragmentation.

For the evaluations in this work we selected three generic cache poisoning methodologies developed in \cite{birge2018bamboozling,cns:frag:dns,saddns}, which are not specific to implementation or setup and do not result due to bugs in randomness generation, such as \cite{klein2020cross}. We perform Internet-wide measurements of these methodologies testing experimentally DNS cache poisoning against DNS resolution platforms. We then exploit these poisoned caches to attack applications that use the poisoned records we injected.

\subsection{DNS Cache Poisoning in the Wild}
There is numerous evidence of DNS cache poisoning attempts in the wild, \cite{dns:venezuela,cmu:mx,myetherwallet,iran,dns-trendmicro-malaysia,dns-fireeye-middleeast,dns-krebsonsecurity-webnic,attack-dnspionage,attack-sea-turtle-returns,attack-liquid}, which were predominantly launched via short-lived BGP (Border Gateway Protocol) prefix hijacks or by compromising a registrar or a nameserver of the domain.

We consider only attacks done by network attackers by manipulating the protocols remotely but without compromising services or networks. Hence compromises of registrars or servers is not in our scope and in the review of works we focus only on BGP prefix hijacks, side channels and fragmentation attacks.

In 2017 \cite{apostolaki2017hijacking} simulated the effects of BGP prefix hijacks on bitcoin without experimentally evaluating it in the wild. In 2018, \cite{birge2018bamboozling} experimentally evaluated the impact of BGP prefix hijacks on domain validation and \cite{brandt2018domain} evaluated the impact of DNS cache poisoning on domain validation. In 2020 a recent research project \cite{DBLP:journals/cacm/SunABVRCM21} evaluated BGP prefix hijacks for cross-layer attacks on Tor (the onion routing) \cite{dingledine2004tor} users, domain validation and bitcoin \cite{franco2014understanding}.

Except for fragmentation based DNS cache poisoning against domain validation \cite{brandt2018domain} there were no studies of cache poisoning using different methodologies and their evaluation against applications. In this work we perform the first comprehensive study of DNS cache poisoning against different applications, and using different methodologies.

\section{Taxonomy of Poisoning Methods}\label{sc:dns:poison}\label{sc:setup}

In our evaluations in subsequent sections we use three methodologies for poisoning DNS caches, which were shown to be practical in previous research: (1) intercepting DNS requests with BGP prefix hijacking \cite{DBLP:journals/cacm/SunABVRCM21}, (2) guessing challenge values in DNS requests via side-channel \cite{saddns} or (3) injecting content into IP defragmentation cache \cite{cns:frag:dns}.
In this section we describe these attack methodologies, their unique properties and explain what attacker capabilities they assume. We compare effectiveness and stealthiness of each of these methods for carrying out cache poisoning attacks. %

{\bf Setup.} To test our attacks experimentally in the Internet we setup a victim AS and associate a /22 prefix with our AS. We register victim domains and setup nameservers and a DNS resolver.

\subsection{Intercepting DNS with BGP Hijacking}\label{sc:hijack}

A malicious Autonomous System (AS) can exploit vulnerabilities in BGP to hijack packets of some victim AS. A route hijack happens when an attacker announces an incorrect prefix belonging to a different AS. 
The attacker hijacks the prefix or a sub-prefix which has the IP address of a DNS nameserver or a resolver. If the hijack succeeds, the ASes that accepted the hijack will send all their traffic destined to the victim prefix instead to the attacker. The goal of the attacker is to intercept a single DNS packet, either a query sent by the resolver or a corresponding response of the nameserver. For simplicity in this discussion we focus on sub-prefix hijacks and assume that the attacker attempts to hijack the DNS query; see \cite{birge2018bamboozling} for a taxonomy of BGP prefix hijack attacks. The attacker intercepts the DNS query and crafts a spoofed DNS response with malicious records and the correct challenge values, and sends it to the victim DNS resolver. Additionally, to avoid detection due to blackholing, the attacker should relay all the traffic to the legitimate destination, except for the DNS query which it intercepted (to avoid race condition with the response from the genuine nameserver). We call this DNS cache poisoning attack method \hijack.

\subsection{Guessing Challenges with Side-channel}\label{sc:saddns} \label{sc:saddns:eval}
The \sad\ off-path attack \cite{saddns} uses an ICMP side channel to guess the UDP source port selected by the victim resolver in its query to the target nameserver. This is done via a side-channel present in most modern operating systems which allows the attacker to test if a given UDP port open or not. The operating systems have a constant, global limit of how many ICMP port unreachable messages they will return when packets are received at closed UDP ports (50 in the case of linux). The attacker splits the range of ports to sets of N ports and for every set performs the following: the attacker sends 50 probes with a spoofed source IP address of the nameserver to a range of UDP ports at the resolver. If the probes arrived at closed ports only, the returned ICMP error messages reach the global limit, and further messages will not be issued. The attacker sends a single probe from the IP address of the attacker to a known-closed port. If all of the previously probed 50 ports were closed the attacker will not receive an ICMP message in response to his own message. However, if one of the 50 probed ports was open, the limit was not reached, the attacker will receive a ICMP port unreachable message. The attacker repeats this process until a set containing an open port is found. Once a set with an open port is found, the attacker applies divide and conquer with the technique above dividing the ports until a single open port is isolated.
This reduces the entropy of the challenge-response parameters unknown to the attacker from 32 bit (DNS TXID + UDP port number) to 16 bit. 

\begin{figure}[t!]
    \centering
    \includegraphics[width=0.47\textwidth]{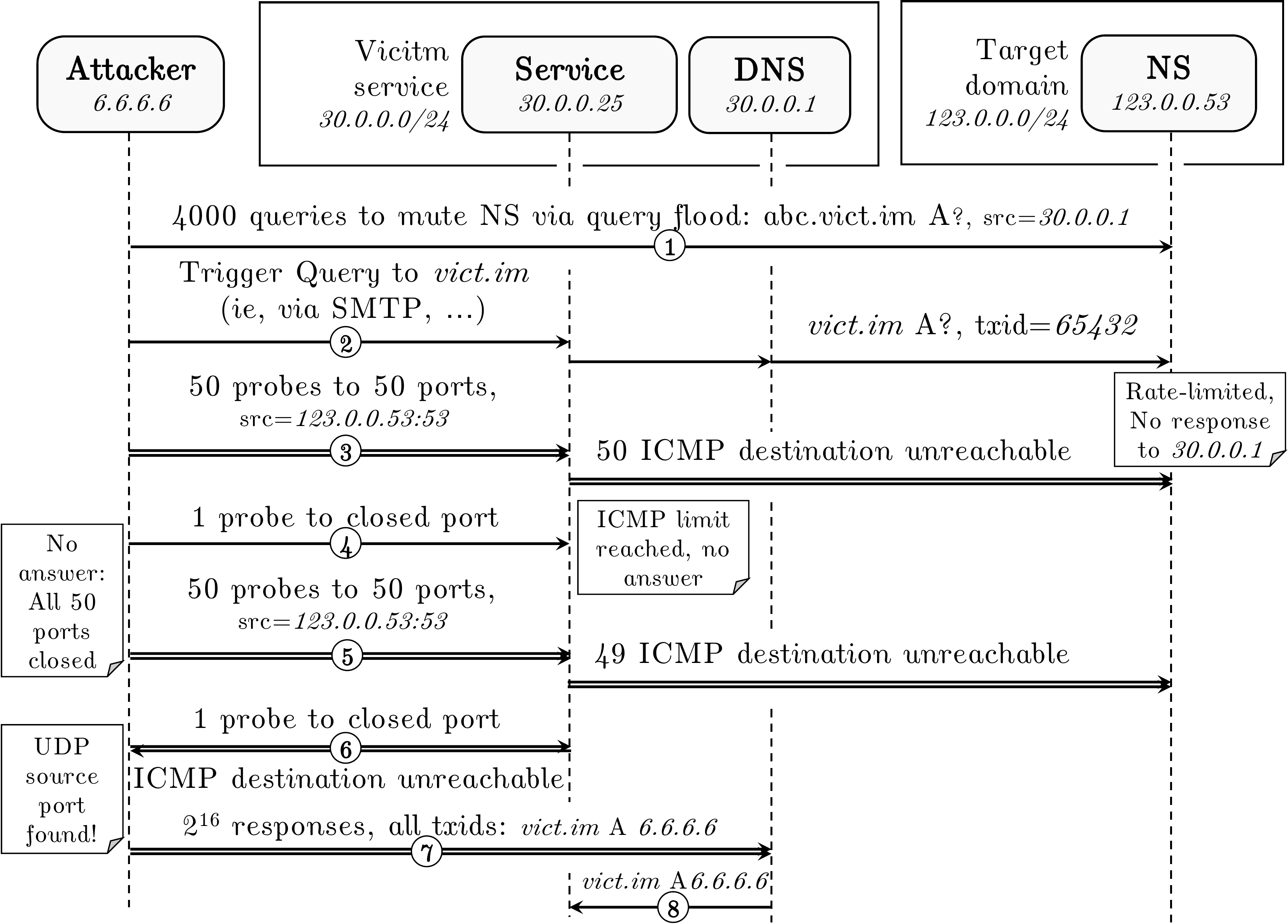}
    \caption{DNS Poisoning with side-channel.}
    \label{fig:saddns_poisoning}
\end{figure}

Once the open port is identified the attacker sends multiple spoofed DNS responses from a spoofed IP address (of the nameserver) to that open UDP port of the resolver, for each possible TXID value, total of $2^{16}$ spoofed responses; e.g., \cite{conf/esorics/HerzbergS12,jeitner2020impact,saddns}. A packet with the correct TXID is accepted by the DNS resolver.
The attack is illustrated in Figure~\ref{fig:saddns_poisoning}.

The attack applies to only about 18\% of the domains with nameservers that use rate-limiting. The rate limiting allows the adversary to delay the response from the genuine nameserver and hence to win the race against it. Additionally, the attack applies only against resolvers with a global (un-patched) ICMP rate limit. 

\subsection{Injecting Records via IP Fragmentation}\label{ssc:frag}
In this section we describe an attack which exploits IP fragmentation to inject spoofed fragments into the IP defragmentation cache on the victim system. The spoofed fragments contain malicious content, which when reassembled with the genuine fragments, manipulate the payload of the original IP packet without having to guess the values in the challenge-response parameters, \cite{cns:frag:dns}. 
\begin{figure}[t!]
    \centering
    \includegraphics[width=0.47\textwidth]{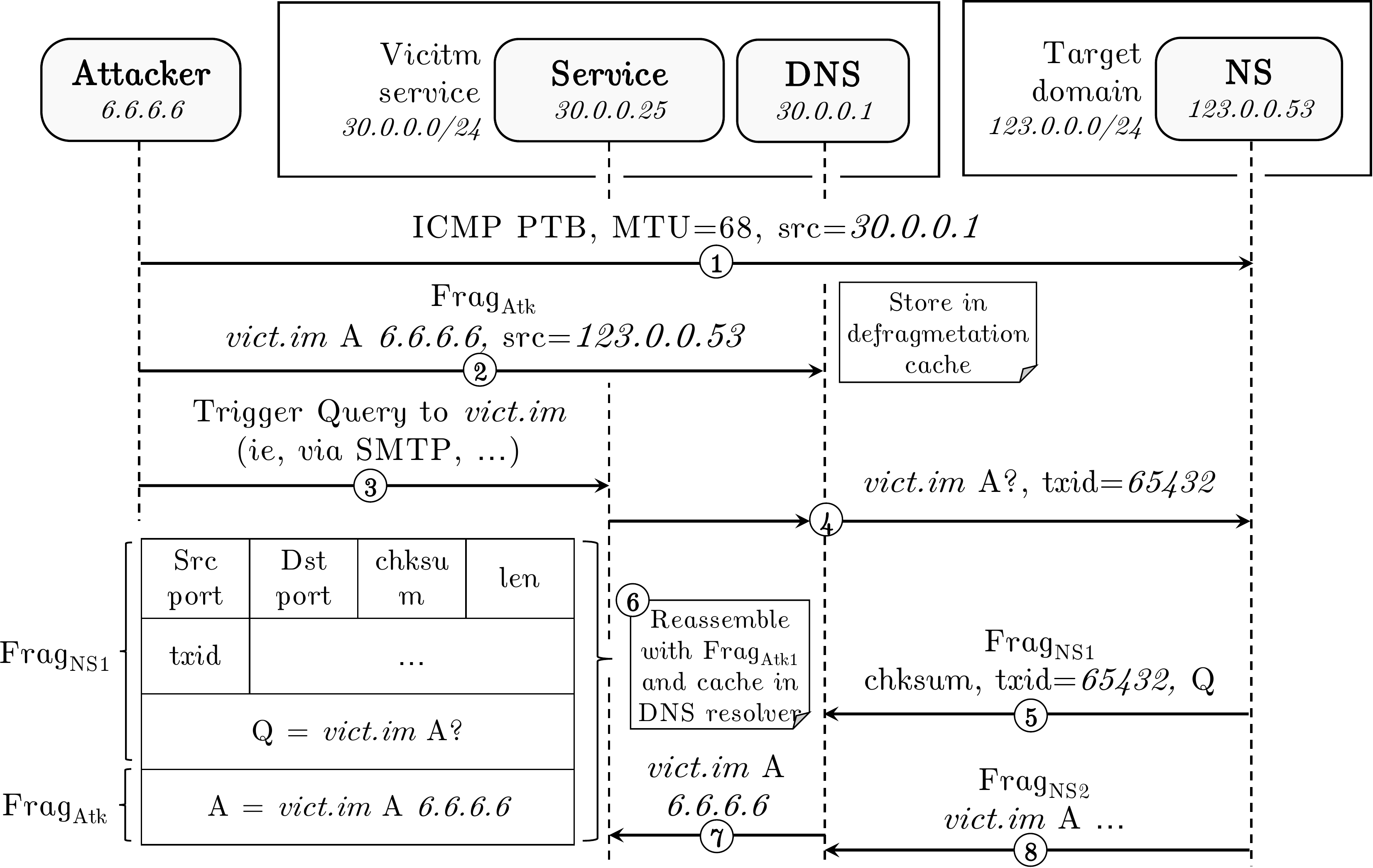}
    \caption{Fragmentation-based DNS poisoning.}
    \label{fig:fragementation_poisoning}
\end{figure}
  
  We assume that the response from the nameserver is fragmented and arrives in at least two fragments. The fragment sent by the attacker is reassembled with the first fragment sent by the nameserver. The attacker replaces the second fragment of the nameserver with its malicious fragment, which overwrites part of the payload of the genuine DNS response from the nameserver, with malicious values. %
  Since the challenge-response values (port, TXID) are in the first fragment, they remain unchanged. The illustration of the attack is in Figure \ref{fig:fragementation_poisoning}.

To cause the nameserver to fragment a DNS response the attacker sends to the nameserver a \textit{ICMP Destination Unreachable Fragmentation Needed} error message (type 3, code 4) with a DF bit set, signalling to the nameserver that the Maximum Transmission Unit (MTU) to the destination is smaller than the packet's length. The nameserver reduces the size of the packet accordingly by fragmenting the IP packet to smaller fragments.

\section{Exploiting DNS Poisoning for Cross-Layer Attacks}\label{sc:taxonomy}\label{sc.taxonomy}
\begin{table*}[ht!]

    \newcommand{\thrpt}{\cmark$^2$}

    \centering
    \footnotesize
    \setlength\tabcolsep{1.5pt}
    \begin{tabular}{|c|c|cH| |rc|c|c| |ccc| |ccc| |HHl|}
    \hline
    \mr{2}{Category}           & \mr{2}{Protocol}         & \mr{2}{Use Case}                        & Section & \hspace{-2pt}query & \mr{2}{\hspace{-1pt}known\hspace{-2pt}} & query trigger                 & Record      & \multicolumn{3}{c||}{ DNS used for } & \multicolumn{3}{c||}{ Methodologies } & Additional & \multicolumn{2}{c|}{ \mr{2}{\makecell{Cache Poisoning\\impact}} } \\ 
       &  &                 &         & \hspace{-2pt}name & & method         & Type        & \hspace{-1pt}loc.\hspace{-1pt}     & \hspace{-1pt}fed.\hspace{-3pt}        & \hspace{-1pt}auth.\hspace{-3pt} & Hijack & \hspace{-1pt}SadDNS\hspace{-1pt} & Frag & Protection*      & vulnerable &  \\
    \hline
    \hline

    \hline
    \multirow{1}{*}{ Authentication }
               & Radius   & Peer discovery  & \ref{ap:sc:radius} & target & \cmark$^1$ & direct          & NAPTR, SRV, A & \cmark   & \cmark      &       & \cmark & \cmark & \cmark  & TLS& (\cmark)   & DoS: no network access \\
    \hline

    \multirow{1}{*}{ Online Chat }
               & XMPP     & Chat+VoIP & \ref{sc:telephony} & target & \cmark$^1$ & bounce           & A, SRV      & \cmark   & \cmark      &       & \cmark & \cmark  & \cmark  &                  & \cmark     & Hijack: eavesdropping \\
               
    \hline
    
    \multirow{3}{*}{ Email } 
               & SMTP     & Mail & \ref{sc:smtp} & target & \cmark$^1$ & direct/bounce     & A, MX       & \cmark   & \cmark      &       & \cmark & \cmark & \cmark &                  & \cmark     & Hijack: eavesdropping \\
               &SPF,DMARC& Anti-Spam & \ref{sc:smtp}              & target & \cmark$^1$ & authentication         & TXT         &          &             & \cmark & \cmark & \cmark & \cmark &                  & \cmark     & Downgrade: spoofing  \\
               & DKIM     & Integrity Checking & \ref{sc:smtp} & target & \cmark$^1$ & direct/bounce     & TXT         &          &             & \cmark & \cmark & \cmark & \cmark &                  & \cmark     & Downgrade: spoofing \\
    
    \hline
    
    \multirow{2}{*}{ Web }
               & HTTP     & Web sites & \ref{sec:web}              & target & \cmark$^1$ & direct      & A           & \cmark   &             &       & \cmark & \cmark & \cmark & (TLS)            & \cmark     & Hijack: eavesdropping \\
               & SMTP     & Password recovery  & \ref{sc:password}      & target & \cmark$^1$ & direct      & A, MX, TXT  & \cmark   &             &       & \cmark & \cmark & \cmark & (2-FA)           & \cmark     & Hijack: account hijack \\
    
    \hline
    
    Sync       & NTP      & Time synchronisation & \ref{ap:sc:ntp}   & known & \cmark & connection DoS         & A           & \cmark   &             &       & \cmark & \xmark & \thrpt &                  & \cmark     & Hijack: change time \\
    
    \hline
    
    \multirow{1}{*}{ \hspace{-1pt}Crypto-currency\hspace{-1pt} }
               & Bitcoin & Peer discovery & \ref{sc:digicash}          & known & \cmark & waiting                & A           & \cmark   &             &       & \cmark & \xmark & \xmark & (fallback)        & (\cmark)   & Hijack: fake blockchain \\
               
    \hline
    
    \multirow{3}{*}{ Tunnelling }
               & OpenVPN   & VPN & \ref{sc:tunnel_normal}                   & config & \xmark & connection DoS         & A          & \cmark    &             &         & \cmark & \thrpt & \thrpt & psk, cert       & \cmark     & DoS: no VPN aceess \\
               & IKE     & VPN & \ref{sc:tunnel_normal}                     & config & \xmark   & connection DoS         & A          & \cmark    &             &         & \cmark & \thrpt & \thrpt & psk, cert       & \cmark     & DoS: no VPN aceess \\
               & IKE     & Opportunistic Enc. & \ref{sec:oppotunistic_ipsec}      & target & \cmark$^1$ & bounce      & IPSECKEY   & \cmark     &             & \cmark  & \cmark & \thrpt & \thrpt & (DNSSEC)        & \cmark     & Hijack: eavesdropping \\
    
    \hline
    
    \multirow{3}{*}{\makecell{ PKI }}
               & DV      & Domain Validation & \ref{sc:domainvalidation}       & target & \cmark$^1$ & authentication         & A, MX, TXT  & \cmark   &             &\cmark & \cmark & \xmark & \xmark &                  & \cmark     & Hijack: fraud. certificate \\
               & OCSP      & Revocation checking & \ref{sec:web}   & target & \cmark$^1$ & direct      & A           &          &             &\cmark & \cmark & \cmark & \cmark & cert             & \cmark     & Downgrade: no check \\
    
               & RPKI      & Repository sync. & \ref{ap:sc:rpki}      & known & \cmark & waiting                & A           & \cmark   &             &       & \cmark & \xmark & \xmark & ROAs        & \cmark     & Downgrade: no ROV \\

    \hline

    \multirow{5}{*}{\makecell{ Intermediate \\ devices }}
               & --       & Firewall filters & \ref{sc:firewalls}       & config & \xmark & waiting                & A           & \cmark   &             &       & \cmark & \thrpt & \thrpt &                  & \cmark     & Downgrade: no filters  \\
               & HTTP/... & Loadbalancers & \ref{sc:cdn}          & config & \xmark & on-demand    & A           & \cmark   &             &       & \cmark & \thrpt & \thrpt & (TLS)            & \cmark     & Hijack: eavesdropping \\
               & HTTP     & CDN's & \ref{sc:cdn}                   & config & \xmark & on-demand    & A           & \cmark   &             &       & \cmark & \xmark & \thrpt & (TLS)            & \cmark     & Hijack: eavesdropping \\
               & DNS      & ANAME/ALIAS\cite{I-D.ietf-dnsop-aname} & \ref{sc:aname}
                                                    & config & \xmark & on-demand    & A           & \cmark   &             &       & \cmark & \thrpt & \thrpt &                  & \cmark     & Hijack: eavesdropping \\
               & HTTP/Socks & Proxies &               & target & \cmark$^1$ & direct      & A           & \cmark   &             &       & \cmark & \cmark & \cmark & (TLS)            & \cmark     & Hijack: eavesdropping \\

    \hline
    
    \end{tabular}
    
    {$^1$:~Depends on the attack scenario. $^2$:~Requires a third-party application to trigger queries.}
    \caption{Evaluation of attacks against popular systems leveraging a poisoned DNS cache.} %
    \label{tab:dns_systems}
\end{table*}
In this section we demonstrate how DNS cache poisoning can be used to launch cross-layer attacks against popular applications. In Section \ref{ssc:method:apps} we explain our methodology for selecting the applications. We list the categories according to which we selected the applications in Table \ref{tab:dns_systems}.
Our analysis of the applications is performed according to the key properties related to cache poisoning: (1) control over the query, (2) which records can be injected, (3) how the application uses the injected records, and (4) the outcome of the attack.

\subsection{Methodology for Selecting Applications}\label{ssc:method:apps}
We select the applications according to the following considerations: application category, usage of DNS by the application and the impact of DNS cache poisoning on the application.

\subsubsection{Category} We categorise the applications to groups, covering most of the popular applications and security mechanisms in the Internet (left most column in Table \ref{tab:dns_systems}). Within each category we selected a few representative protocols and systems for that category, see column `Protocol' in Table \ref{tab:dns_systems}. 

\subsubsection{Usage of DNS} One of the considerations for selecting the applications is how the application uses DNS: how the queries are sent by the application to the DNS resolver and how the results from the lookups are processed. The column `Use-Case' in Table \ref{tab:dns_systems} describes the usage scenarios of the DNS by the application. We defined the following types:

\textit{Location (loc):} DNS is used to locate a direct communication partner, typically in form of a hostname-to-ip (\texttt{A, AAAA}) mapping.

\textit{Federation (fed):} DNS is used to locate a user's home-server based on the domain part of a user address of the form \texttt{user@domain}.

\textit{Authorisation (auth):} DNS is used to authorise a certain action or host in the name of the domain's owner.

\subsubsection{Queries} Applications differ in flexibility in allowing external entities to trigger queries. Our selection of applications aims to cover the variety of options for triggering queries. To initiate the attack, our adversary needs to cause the victim resolver to issue the target query or to predict when the query will have been issued. %

Some applications enable the attacker to send arbitrary queries, e.g., in systems which use DNS for peer discovery in federated systems like Radius, XMPP and SMTP. This is because in these systems, the queried domains are part of the user's ID. This user ID can be controlled by the attacker to trigger a query to a domain of its choice. The same applies to all (sub-)systems used as part of web browsing, like HTTP, DANE and OCSP, since the attacker can establish direct connection from the victim client to arbitrary web servers which will trigger a DNS lookup that way. Setting the domain name is not always possible, e.g., in NTP the query is selected by the resolver based on the hostname that it receives from the local NTP server.%

\begin{added}
We evaluated popular appliances and systems for their query triggering behaviour. We list some selected systems in Table \ref{tab:cdn_dns_resolver_test}. As can be seen, some allow external adversaries to trigger queries (indicated with "on-demand" in column Trigger query). Other devices use timers for issuing queries. Hence the adversaries can often predict when the query is issued.
\end{added}

\subsubsection{Impact of poisoning on applications}
 We select applications to demonstrate the impact that cache poisoning on applications can create: DoS (Denial of Service), downgrade of security or interception attacks. %

\begin{table}
    \centering
    \small
    \setlength\tabcolsep{3.4pt}
    \begin{tabular}{|c|r|c|cHH|r|}
    \hline
    \mr{2}{Type}
    &\mr{2}{Provider}         & Trigger & Caching & Accepting  & IPs* & Websites in    \\
    &                 & query     & time          & Fragments  &      & Alexa 100K     \\ %
    \hline
    \hline
    \mr{2}{\makecell{Firewall}}
    &pfSense         & timer    & 500s         & -          & -  &    - \\ %
    &Sophos UTM      & timer    & 240s         & -          & -  &  - \\ %
    \hline
    \hline
    \mr{2}{\makecell{Load\\balancer}}
    &Kemp Technologies& timer    & 1h           & \cmark     & -  &    - \\ %
    &F5 Networks      & timer    & 1h           & -          & -  &  - \\ %
    \hline
    \hline
    \mr{4}{\makecell{CDN}}
    &Stackpath        & on-demand& TTL          & \cmark     & 2    &    79 \\ %
    &Fastly           & timer    & TTL          & -          & 180  &  1,143 \\ %
    &AWS              & on-demand& TTL          & -          & 6    & 11,057 \\ %
    &Cloudflare       & on-demand& TTL          & -          & 4    & 17,393 \\ %
    \hline
    \hline
    \mr{4}{\makecell{Managed\\DNS\\(ALIAS)}}
    &DNSimple         & on-demand& TTL          & \cmark     & 7    &   248 \\ %
    &DNS Made Easy    & timer    & $\sim$35min  & -          & 4    &  1,192 \\ %
    &Oracle Cloud     & on-demand& TTL          & \cmark     & 21   &  1,382 \\ %
    &Cloudflare       & on-demand& TTL          & -          & 3    & 20,027 \\ %
    \hline

    \end{tabular}
    \caption{Query triggering behaviour at middleboxes. Last column shows the number of websites in 100K-top Alexa which use that provider.}%
    \label{tab:cdn_dns_resolver_test}
\vspace{-10pt}
\end{table}

\begin{added}

\subsection{Methodology for Attacking Applications}\label{sc:attacks:apps}

We developed cross-layer attacks that leverage DNS cache poisoning to attack applications listed in Table \ref{tab:dns_systems}. The steps underlying all our cross-layer attacks against applications are: 

(1) Use the application to send to the victim DNS resolver a query. In addition to the traditional ways of triggering queries, such as with a script or Email, we also developed new ways to trigger queries which were not known prior to our work. Some of these techniques are specific to appliances and platforms, see Table \ref{tab:cdn_dns_resolver_test}, while others are application-independent methodologies for triggering queries. We explain our methodologies for triggering queries in Section \ref{sc:trigger:queries}. 

(2) Inject malicious records to poison the cache of the victim DNS resolver. We use the methodologies in Section \ref{sc:dns:poison} for injecting malicious records into the cache of the victim DNS resolver. In Table \ref{tab:dns_systems} we summarise the applicability of the cache poisoning methodologies for cross-layer attacks against each application, and explain this in Section \ref{sc:applicability}.

(3) Exploit the poisoned records to cause a victim application to divert from the expected behaviour.  %
The outcomes of our cross-layer attacks against applications range from downgrading security to denial of service attacks and to more traditional impersonation attacks, explained in Section \ref{sc:exploits:records}.

\end{added}
\subsection{Methodologies for Triggering Queries}\label{sc:trigger:queries}

\begin{added}
\subsubsection{Common ways for triggering queries}
The most challenging aspect of cross layer attacks that use DNS cache poisoning is the ability to trigger or predict DNS requests. Typically an external adversary does not have access to internal services, such as the DNS resolver, and hence should not be able to cause the DNS resolver to issue arbitrary DNS requests. %
Adversaries can trigger queries via bounce. For instance, by sending an Email to a non existing recipient in the target domain the adversary will cause the Email server to return an error message with Delivery Status Notification. To send the error the Email server requires the IP address and hostname of the MX server in the domain that sent the Email message which triggered the error. This causes queries to the domain specified by the attacker. 

The adversary can also set up a web server and lure clients to access it, this is a direct query triggering. %
The clients download the web objects from the adversary's domain, and send DNS requests to the DNS resolvers on their networks. When resolvers receive DNS requests from servers or clients on their networks they initiate DNS resolution. However, these approaches are limited. For instance, only about 18\% of the Email servers trigger DNS requests when receiving Emails, \cite{klein2017internet}. The limitation with web clients is that the adversary must wait until the target client visits the malicious web page. Furthermore, web clients cannot be used to poison resolvers that are used only by servers, such as Email or NTP. In this section we develop new approaches for triggering queries. %
\end{added}

\subsubsection{Cross-applications DNS caches}
The adversary may be able to use one application to trigger queries to inject a record that is meant to be used for cross-layer attack against a different application that uses the same DNS cache. For instance, when an adversary cannot trigger queries via an application that it wishes to attack, it may often be able to trigger queries via a different application, that uses the same DNS cache. The adversary may also choose to inject into such cross-applications caches an application agnostic records; for instance, a malicious NS record, mapping the nameserver of a domain of the target application to the attacker's IP address, is an example of an application agnostic record. 

Such cross-applications DNS cache scenario is not uncommon. The DNS resolvers often serve multiple applications and the networks use the caching of the resolver to reduce traffic and latency for all the applications.  %
We use open resolvers to check how common cross-applications DNS caches are. We perform our measurements against a list of open resolvers from censys \cite{censys15} and probe their caches for the well-known domain(s) used by the applications on our list in Table \ref{tab:dns_systems}, e.g., {\tt pool.ntp.org} for NTP. For each application for which the records are in the cache we consider that the resolver is used by that application or by the clients of that application. We found that 69\% of the open resolvers are shared between two or more of the applications on our list. 

A recent study \cite{jeitner2020impact} analysed how an attacker can find third-party SMTP servers to trigger queries at typically closed forwarders used by web clients. By scanning the /24 network block of the resolver's outbound IP address, the study found that an adversary could find an SMTP server which allows triggering queries from the same resolver in 11.3\% of the cases. Additionally 2.3\% of the resolvers were open resolvers in the first place.

\subsubsection{Triggering queries via forwarders}
\label{sec:third_party_triggering}

In this section we show how to trigger queries with resolvers when this is not possible from the target application. %

DNS forwarders make up the majority of open resolvers in the internet. Finding an open forwarder which forwards to the resolver of choice whose cache the adversary wishes to poison is not difficult. We explore the prevalence of forwarders through which one can force a given recursive DNS resolver to trigger a query. We perform a two step measurement: we first collect the forwarders used by open DNS resolvers and then which of these forwarders are used by random clients in the Internet. %

In our measurement we use the list of all open DNS resolvers from Censys \cite{censys15} (which performs a full IPv4 scan for open resolvers). We query all the resolvers for a custom query with a randomised subdomain under a domain which we control. This allows us upon the arrival of the DNS requests to our nameservers to map the open resolver's IP address to the recursive forwarder that it uses. This forwarder is determined by the outbound IP address in the DNS query that arrives at our nameserver. 

In the second step we run a web ad-based study against random clients in the Internet that download our object. We trigger DNS requests via those clients to our own domain. We use a random subdomain associated with each client. Per client, we then obtain the list of recursive resolvers' IP addresses that arrived at our nameserver. We search them in the list of recursive resolvers IP addresses from our dataset of open resolvers. 

Our results are as follows: focusing only on the IP addresses of the recursive resolvers, we find 4146 addresses out of which 3275 (79\%) addresses are in the open resolver database. \begin{added} Consequently, assuming that an adversary targets a DNS resolver used by a typical web client (represented by the ad-net clients in our study), there is a high probability (79\%) that it can find an open forwarder which can be used to poison the cache of the target victim recursive resolver used by that web client. \end{added} %

\begin{added}
\subsection{Applicability to Applications}\label{sc:applicability}
In this section we explore which cache poisoning methodology is applicable to which of the applications listed in Table~\ref{tab:dns_systems}. %

For all methodologies, the attacker requires the knowledge of the domain which is queried. In cases where the domain is pre-configured in the applications configuration ("config" in Table~\ref{tab:dns_systems}), this information needs to be fetched out of band. %

\subsubsection{\hijack{}}
The adversary can hijack a sub-prefix or same-prefix of the victim AS. We explain the success probability of cache poisoning through both methods.

\textbf{Sub-prefix hijack.} The attacker can advertise a sub-prefix of the victim. The routers prefer more specific IP prefixes over less specific ones, hence this announcement will redirect all traffic for that sub-prefix to the attacker. %

\textbf{Same-prefix hijack.} Same-prefix hijack occurs when the attacker hijacks a route to an existing IP prefix of the victim. The attacker can advertise the same prefix as the victim AS and depending on the local preferences of the ASes will intercept traffic from all the ASes that have less hops (shorter AS-PATH) to the attacker than to the victim AS. The success of the hijack depends on the topological relationship between the attacking AS and the domain and the victim resolver. %

\subsubsection{\sad{}} The attack is probabilistic since it depends on the ability of the adversary to win the race, by correctly guessing the randomised TXID before the timeout event. A prerequisite to a successful attack is the ability to trigger a large volume of queries. Typically, this is the case when the query domain can be set by the attacker ("target" in Table~\ref{tab:dns_systems}, Column "query name") or when a third party application is used to trigger the queries (marked with \cmark{}$^2$ in Table~\ref{tab:dns_systems}, see Section~\ref{sec:third_party_triggering}).

\subsubsection{\frag{}}
\frag{} is also a probabilistic attack since its success depends on correctly guessing the IP ID value in the spoofed IP fragment. This is easy when systems have large IP defragmentation buffers, such as old linux versions which allows the adversary to send multiple fragments with different IP ID values, or when systems use incremental IP ID counters which can be predicted. A successful poisoning with \frag{} typically requires more packets than with prefix hijacks but less than with \sad{} attack.

\end{added}

\subsection{Exploiting Poisoned Caches for Attacks}\label{sc:exploits:records}
Applications that use DNS resolvers with poisoned caches are exposed to a range of attacks. In this section we explain the possible outcomes of the attacks via DNS cache poisoning.

{\bf Downgrade attacks.} In downgrade attacks the attacker makes the security mechanism not available, as a result, causing the processing of the data to be performed without the additional information provided by the security mechanism. For instance, by poisoning the responses to queries for SPF or DKIM records the attacker can trick the victim Email server into accepting phishing Emails or Emails with malicious attachments. Similarly, by causing the RPKI validation to fail, the adversary can make a network, that filters bogus BGP announcements with route origin validation, to accept hijacked prefixes as authentic. This is due to the fact that RPKI validation will result in status `unknown' and hence will not be used. %

The attacker can also trick a security mechanism via DNS cache poisoning. For instance, the attacker can bypass domain validation, by redirecting the validation to run against attacker's host \cite{brandt2018domain}, and hence can issue fraudulent certificates.

{\bf Hijack attacks.} In hijack attacks the victims are redirected to attacker's host which impersonates a genuine service in the Internet. Network adversaries can hijack traffic to take over Internet resources, such as SSO accounts at public providers. For instance, the adversaries can take over the SSO accounts at Regional Internet Registries (RIRs), by exploiting a combination of DNS cache poisoning with password recovery \cite{274630}. The idea is to poison the cache of the RIR, and to inject a record that maps the victim LIR to the host of the attacker. Running a password recovery procedure causes the password for the victim SSO account to be sent to the attacker instead of the victim. As a result, the attacker can hijack the digital resources, such as IP addresses and domains, that belong to the victim LIR.

{\bf DoS attacks.} The attacker can block connectivity, e.g., for radius clients or access to services, such as secure tunnels. The idea is that if the attacker cannot forge cryptographic material, such as a certificate to authenticate a radius client, it can redirect the client to the wrong host via cache poisoning, preventing the client from connecting to the genuine target service. The adversary will not be able to provide authenticated material which will result in a failure, and lack of service for the victim client.

\section{Internet Measurements}\label{sc:eval}

In this section, we analyse the fraction of the vulnerable resolvers and nameservers with respect to each DNS poisoning method. We evaluate properties which influence the success of the cross-layer attacks against applications.\begin{added} Our measurements in this section show that the vulnerabilities do not significantly differ for most of the application-specific datasets. The outliers can be summarised as follows:

$\bullet$ Vulnerabilities to BGP sub-prefix hijacking are exceptionally high for eduroam and low for RPKI domains. The cause may be inherent in networks' sizes (large in case of universities and small for RPKI repository operators) and accordingly use BGP announcements which are larger than /24 for large networks or equal to /24 for small networks.

$\bullet$ Vulnerabilities to fragmentation cache poisoning among open resolvers is low compared to other resolvers in our dataset. This may be due to the fact that the distribution of the open resolvers is skewed towards poorly configured devices which cannot handle fragmentation.

$\bullet$ Domains with MX, SRV, NAPTR (eduroam) records are more often vulnerable to fragmentation based cache poisoning than the domains in the 1M-top Alexa dataset. One reason is that the responses to ANY queries result in much larger packets, which often exceed the minimum MTU limit.
\end{added}

\subsection{Vulnerabilities in Resolvers}
\label{sc:measurement-vuln-resolver}

We test the DNS resolvers for vulnerabilities to the three cache poisoning methods (Section \ref{sc:dns:poison}) for different applications. The results of our evaluations for all datasets and all poisoning methods are summarised in Table~\ref{tab:vulnerable_resolvers_summary}.  %

\subsubsection{Dataset}

\begin{added}
For each application from Section~\ref{sc:taxonomy}, we gather datasets of resolvers used by the front-end systems (i.e., Web clients, Alexa MX records, etc.) of that application. To achieve this, we first look for an appropriate dataset of front-end systems and then trigger queries through those front-end systems. This allows us to discover and test the corresponding resolver.

\end{added}

\begin{added}For front-end systems, we use the following\end{added} datasets, listed in Table~\ref{tab:vulnerable_resolvers_summary}: 
(1) Our local university eduroam service. (2) Password recovery of popular infrastructure service providers, consisting of: All 5 Regional Internet Registries, popular domain registrars used by Alexa Top 100K domains and popular cloud providers \cite{gartner-iaas-market-share,srgresearch-iaas-market-share,random-iaas-list-1,random-iaas-list-2}. (3) Domain validation of most popular Certificate authorities \cite{ca-market-share}. (4) Popular CDNs in Alexa Top 100K (by mapping A record to ASN). (5,6) SMTP and XMPP servers of Alexa Top 1M domains. (7) Web clients gathered via an Ad-network. (8) Open resolvers from Censys \cite{censys15} and (9) a subset of those open resolvers who cache \path{pool.ntp.org}.
This resulted in a dataset of 89,924 resolvers (back-end IP addresses) in 13,804 ASes associated with 33,418 prefixes.

\begin{added}
We report the dataset size in terms of front-end systems (i.e., number of SMTP servers or number of open resolver front-end IP addresses) in column "Dataset size" of Table~\ref{tab:vulnerable_resolvers_summary}. For vulnerability, we report the percentage of vulnerable front-end systems which was measured as described in Section~\ref{ssc:res}. When a front-end system uses multiple resolvers, we consider it vulnerable if any of the resolvers it uses is vulnerable.
\end{added}

\subsubsection{Measuring cache poisoning vulnerabilities in resolvers}\label{ssc:res}
The results of our measurements and evaluations against resolvers for different poisoning methodology are summarised in Table \ref{tab:vulnerable_resolvers_summary}. In the following sections we explain the measurements we carried out of each attack methodology against the resolvers in our dataset.

\paragraph{Sub-prefix BGP hijacks (\hijack{})} Since many networks filter BGP advertisements with prefixes more specific than /24, we consider an IP address hijackable if it lies inside a network block whose advertised size is larger than /24. We therefore map all the resolvers' IP addresses to network blocks and consider those vulnerable to sub-prefix hijacks whose advertisement is larger than /24, since an advertisement with a smaller prefix will always take precedence over a bigger one. For the remaining addresses, a BGP-hijack may still be possible using same-prefix hijacks. To infer the scope of DNS platforms potentially vulnerable to cache poisoning via BGP sub-prefix hijack attacks we perform Internet measurements checking for DNS platforms on prefixes less than /24. We collect information on the state of the global BGP table in the Internet with Routeview \cite{routeviews} and RIPE RIS \cite{riperis} collectors. We analyse the BGP announcements seen in public collectors for identifying networks vulnerable to sub-prefix hijacks by studying the advertised prefixes sizes. The measurements of resolvers vulnerable to BGP sub-prefix hijacks are listed in Table \ref{tab:vulnerable_resolvers_summary} and plotted in Figure \ref{fig:prefix_lengths}.

\paragraph{Same-prefix BGP hijacks (\hijack{})} We perform simulations of same-prefix BGP hijacks using a set of randomly selected attacker and victim AS pairs using a simulator developed in \cite{hlavacek2020disco} and Internet AS level topology downloaded from CAIDA \cite{caida11}. The simulator selects Gao-Rexford policy compliant paths \cite{gao2001stable}, and considers prefix lengths and AS-relationship (provider, customer and peer) and sizes (stub, small, medium and tier one).
The attackers are randomly selected from all the ASes whereby the victim ASes are selected from our dataset of DNS resolvers and 1M-top Alexa domains. For each (attacker,~victim)-pair we perform a simulation of same-prefix hijack that the attacker AS launches against a victim AS. If the attacking AS is closer to the victim, the attack succeeds. The simulation shows that the attacking AS was capable of hijacking the traffic in 80\% of the evaluations.

\paragraph{\sad{}}
To test resolvers vulnerable to \sad, we test the resolvers back-end IP addresses for a global ICMP message limit which allows to use the side-channel identified by \cite{saddns}. %
To limit our dataset to functional resolvers which are still reachable, we furthermore send an ICMP echo-response (`ping`) packet to the resolver first. This is especially important for the open resolvers dataset, since this dataset tends to include resolvers operating from dynamic IP addresses, which may have changed since the dataset was collected.

\begin{added}
For the open resolver dataset we measured a vulnerability rate of 12\%, a notable reduction from the 35\% vulnerability rate of the original paper \cite{saddns}. This difference could be influenced by various factors including the fact that our dataset contained more resolvers than \cite{saddns}. The crucial difference is likely that our study was conducted after the vulnerability which allowed the global rate limit to be exploited was patched in many systems. For example, all updated versions of Ubuntu should have been patched by the time we carried out our evaluations\footnote{\url{https://ubuntu.com/security/CVE-2020-25705}}.
\end{added}

\paragraph{\frag{}}
To test vulnerability to fragmentation-based DNS cache poisoning, we use a custom nameserver application which will always emit fragmented responses padded to a certain size to reach the tested fragment size limit. The nameserver is configured to only send CNAME responses in the first fragmented response. This means that if the resolver receives a fragmented response, it needs to re-query for the CNAME-alias. This allows us to verify that the answer arrived at the resolver and thus, that the resolver is vulnerable to this type of attack.

\begin{added}
Using this setup, we test all resolvers by triggering queries to our nameservers and observe if the fragmented responses are accepted.
In our bigger datasets, vulnerability rates range between 31\% for Open resolvers and 91\% for Ad-net resolvers.
For the smaller datasets, we still observe many vulnerable services. However, all certificate authorities' resolvers in our dataset rejected our fragmented responses, maybe attributed to the fact that this attack method was already evaluated and disclosed to CAs previously \cite{brandt2018domain}.
We report results for all datasets and all poisoning methods in Table~\ref{tab:vulnerable_resolvers_summary}. 
\end{added}

\begin{table}
    \small
    \centering
    \setlength\tabcolsep{3.4pt}
    \begin{tabular}{|lH|c|cH|c|c|r|}
\hline

\mr{3}{Dataset} & Category & \mr{3}{Protocol} & \mcc{4}{Vulnerable against} & \mr{3}{\makecell{Dataset\\size}} \\ %
\cline{4-7}
                  &                       &           & \mcc{2}{BGP hijack} & Sad- & Frag-   &  \\ \cline{4-5}
                  &                       &           & sub-prefix & same- & DNS & ment   &      \\

\hline
\hline

 (1) Local university         & Auth.                 & Radius    & 100\%      &             & 0\%       & 100\%      & 1       \\ \hline 
 (2) Popular services &Service providers&\makecell{PW-\\recovery }& 93\%       &             & 16\%      & 90\%       & 29      \\ \hline  
 (3) Popular CAs              & PKI                   & DV        & 75\%       &             & 0\%       & 0\%        & 5       \\ \hline 
 (4) Popular CDNs             & Interm.               & CDN       & 100\%      &             & 0\%       & 25\%       & 4       \\ \hline 
 \hline
 (5) Alexa 1M SRV             & Chat                  & XMPP      & 73\%       &             & 1\%       & 57\%       & 476     \\ \hline
 (6) Alexa 1M MX  & Email & \makecell{SMTP\\SPF\\DMARC\\DKIM}     & 79\%       &             & 9\%       & 56\%       & 61,036     \\ \hline
 (7) Ad-net study & Web   & \makecell{HTTP\\DANE\\OCSP}           & 70\%       &             & 11\%      & 91\%       & 5,847       \\ \hline
 (8) Open resolvers           & All                   & All       & 74\%       &             & 12\%      & 31\%       & 1,583,045  \\ \hline
 (9) Cache test               & Sync                  & NTP       & 79\%       &             & 9\%       & 32\%       & 448,521    \\ 
\hline

    \end{tabular}
    \caption{Vulnerable resolvers.}
 \vspace{-20pt}
    \label{tab:vulnerable_resolvers_summary}
\end{table}

\subsection{Vulnerabilities in Domains}
\label{sc:measure-ns-vuln}

In this section we perform measurements of the vulnerabilities in domains to our cache poisoning methodologies for different applications. We collect lists of the domains associated with these applications and test all the nameservers serving each domain according to the properties required for each cache poisoning method (from Section \ref{sc:dns:poison}). The results of our evaluations and measurements for all the tested datasets and poisoning methods are summarised in Table~\ref{tab:vulnerable_domains_summary}. 
\begin{table}[ht!]
    \small
    \centering
    \setlength\tabcolsep{1.1pt}
    \begin{tabular}{|l|c|cH|c|c|c|c|r|}
\hline

 \mr{3}{Dataset}          & \mr{3}{Protocol}       & \mcc{5}{Vulnerable against}                      &\mr{3}{\makecell{DNS\\SEC}}& \mr{3}{Total} \\ \cline{3-7}
                          &                        & \mcc{2}{BGP hijack} & Sad-     &\mcc{2}{Fragment}&              &       \\   
                          &                        & sub-prefix     & same    & DNS      & Any    & Global &              &       \\ 
\hline \hline

 (1) Eduroam list         & Radius                 & 96\%     &          & 11\%     & 44\%   &  18\%  & 10\%         & 1,152   \\ \hline
 (2) Alexa 1M & \makecell{HTTP\\DANE\\DV}          & 53\%     & 70\%     & 12\%     &  4\%   &  1\%   & 2\%          & 877,071 \\ \hline %
 
 (3) Alexa 1M MX &\makecell{SMTP\\SPF\\DKIM\\DMARC}& 44\%     &          & 6\%      &  7\%   & 1\%    & 3\%          & 63,726  \\ \hline
 (4) Alexa 1M SRV         & XMPP                   & 44\%     &          & 4\%      & 29\%   & 5\%    & 7\%          & 2,025   \\ \hline

 (5) RIR whois            & PW-                    & 59\%     &          & 9\%      & 14\%   & 4\%    & 4\%          & 58,742  \\ \cline{1-1} \cline{3-9} %
 (6) Registrar whois      & recovery               & 51\%     &          & 10\%     & 23\%   & 5\%    & 6\%          & 4,628   \\ \hline %
 (7) Well-known           & NTP                    & 25\%     &          & 0\%      & 25\%   & 25\%   & 25\%         & 9       \\ \hline
 (8) Well-known    & \makecell{Crypto-\\ currency} & 28\%     &          & 17\%     & 21\%   & 3\%    & 21\%         & 32      \\ \hline
 (9) Well-known           & RPKI                   & 14\%     &          & 0\%      & 0\%    & 0\%    & 67\%         & 8       \\ \hline
(10) Cert. Scan           & \makecell{IKE\\OpenVPN}& 51\%     &          & 11\%     & 5\%    & 1\%    & 7\%          & 307      \\ \hline

    \end{tabular}     
    \caption{Vulnerable domains.}
    \label{tab:vulnerable_domains_summary}
    \vspace{-10pt}
\end{table}

\subsubsection{Dataset}

\begin{added}
For each application in Section~\ref{sc:taxonomy}, we collect datasets of typical domains looked up by clients (or servers) of that application. 
\end{added}
We collect such domains from the following data sources, listed in Table~\ref{tab:vulnerable_domains_summary}:

(1) Eduroam institution lists from United Kingdom \cite{eduroam-list-uk}, Germany \cite{eduroam-list-de} and Austria \cite{eduroam-list-at}.
(2) Alexa Top 1 Million domains, including subsets of domains which have (3) MX and (4) SRV (XMPP) records.
Domains from account email addresses from whois databases of (5) RIRs and (6) Registrars.
(7) Well-known NTP server domains.
(8) Well-known cryptocurrency domains.
(9) Well-known RPKI validator database domains.
(10) Domains of IKE and OpenVPN servers' certificates.
This resulted in 904,555 domains hosted on 200,086 nameservers in 24,353 ASes associated with 60,511 prefixes.

\subsubsection{Measuring cache poisoning vulnerabilities in nameservers}

\paragraph{\hijack{}}

\begin{figure}[t]
    \centering\includegraphics[width=0.98\columnwidth]{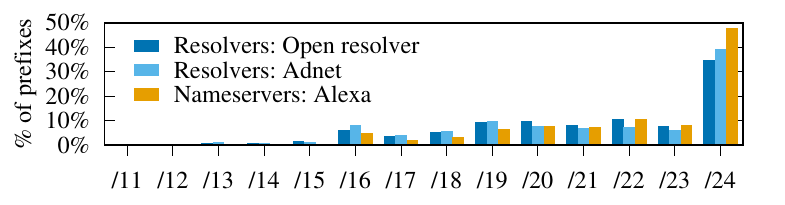}
    \vspace{-10pt}
    \caption{Announced prefixes.}
    \vspace{-10pt}
    \label{fig:prefix_lengths}
\end{figure}

We perform a similar analysis as in Section \ref{ssc:res}, to check the vulnerabilities to  BGP prefix hijacks. The results are plotted in Figure \ref{fig:prefix_lengths}. The differences between the fractions of nameservers in 1M-top Alexa domains that can be sub-prefix hijacked are not significantly different than those of the resolvers. %

The resilience of the DNS infrastructure to BGP hijack attacks is also a function of the distribution and the topological location of the nameservers in the Internet. We measured the characteristics of the nameservers from the Internet routing perspective. Our findings show that the nameservers are concentrated in just a few ASes. Our measurements show that 80\% of the ASes host less than 10\% of the nameservers, and the rest of the nameservers are concentrated on the remaining ASes. This concentration of the nameservers on a few ASes, typically CDNs, makes it easier to intercept traffic of multiple nameservers with a single prefix hijack.

\paragraph{SadDNS}

For a nameserver to be vulnerable to side-channel attack (Section \ref{sc:saddns}), the attacker must be able to `mute` the nameserver to extend the time-window for the attack. This is achieved by abusing rate-limiting in nameservers. To find out if a nameserver supports rate-limiting, we use the \begin{added}following\end{added} methodology: we send to the nameserver a burst of \begin{added}4000\end{added} queries in one second, and see if this stops (or reduces) the subsequent responses received from this server. We consider a nameserver to be vulnerable if we can measure a reduction in responses after the burst.

\paragraph{Fragmentation}\label{sc:ns-fragment-size-any-vs-a}\label{sc:measure-resolver-edns-size}

We evaluate the vulnerability to fragmentation-based poisoning in nameservers and domains by testing three properties required to create a sufficiently large fragment in order to inject malicious records into it: (1) support of IMCP fragmentation needed, (2) record types for optimising response size, (3) by bloating the queried domain and (4) fitting the response into the limitation of EDNS.

{\bf PMTUD.} We first check for the support of path MTU discovery (PMTUD) with ICMP fragmentation needed: we send to the nameserver an ICMP fragmentation needed packet, which indicates that the nameserver should fragment packets sent to our test host. Then we send queries of different type to that domain. We consider a nameserver vulnerable if the responses return fragmented. %

\begin{figure}[t]
    \centering
    \includegraphics[width=0.47\textwidth]{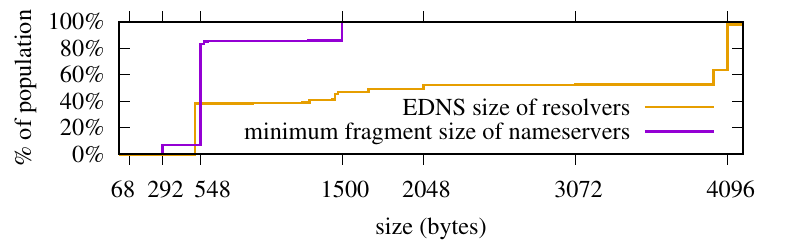}
    \vspace{-10pt}
    \caption{CDF of resolver EDNS UDP size vs. minimum fragment size emitted by nameservers.}
    \label{fig:edns_size}
    \label{fig:alexa_top_1m_domains_fragsize}
\end{figure}

{\bf Record types.} We evaluated fragmentation with three record types: \path{ANY}, \path{A} and \path{MX}. We use DNS requests of type \path{ANY} to increase the response size above the fragmentation limit of the nameserver. We find that for 19.50\% of domains in 1M-top Alexa there is at least one nameserver which emits fragmented DNS responses, which can be used for cache poisoning attacks via injection of IP fragments. We plot the minimum fragment size emitted by those nameservers in Figure \ref{fig:alexa_top_1m_domains_fragsize}, which shows that most affected nameservers (83.2\%) fragment DNS responses down to a size of 548 bytes and 7.05\% even down to 292 bytes. We tested \path{ANY} response caching in 5 of the most popular resolver implementations and found that 3 out of 5 use the contents of an \path{ANY} response, to answer subsequent A queries, without issuing further queries (See Table~\ref{tab:any-caching}). Namely, the adversaries can often launch cache poisoning attacks by issuing queries for \path{ANY} record type in the domain.

However, only open resolvers (or forwarders) allow the attacker to trigger \path{ANY} queries. We repeat the same study using queries for \path{A} record type and then for \path{MX} record type, which are the query types typically triggered using the other query-triggering methods, such as via email or a script in a browser.
We get vulnerability rates of 0.29\% and 0.44\% respectively due to the smaller response sizes which are often not sufficiently large to reach the nameserver's minimum fragment size. However, these numbers represent the lower bound.

{\bf Bloat query.} The attacker can bloat the queries by concatenating multiple subdomains which increases the responses sizes. The maximum increase is up to 255 characters. The labels are limited to max 63 characters (+1 for the label delimiter) and the attacker can concatenate four subdomains: 4*64 (minus the parent domain). This increases the vulnerable resolvers to above 10\%.

\begin{table}
    \small
    \centering
    \begin{tabular}{|r|c|l|}
\hline
Implementation            & Vulnerable & Note \\
\hline
BIND 9.14.0               & yes        & cached \\
Unbound 1.9.1             & no         & doesn't support ANY at all \\
PowerDNS Recursor 4.3.0   & yes        & cached \\
\hline
systemd resolved 245      & yes        & cached \\
dnsmasq-2.79              & no         & not cached \\
\hline
    \end{tabular}
    \caption{ANY caching results of popular resolvers.}
    \label{tab:any-caching}
    \vspace{-20pt}
\end{table}

{\bf Fitting into response.} Additionally to the requirement that the DNS response size must be big enough to trigger fragmentation on the nameserver side, it must also be small enough to fit in the resolvers maximum response size advertised in EDNS.

To evaluate this, we measure the EDNS UDP size of more than 1.5K open resolvers collected from Censys \cite{censys15} IPv4 Internet scans. We query each resolver by triggering a query to our own nameserver and measure the EDNS UDP payload size advertised in the query. The results are shown in Figure~\ref{fig:edns_size}. Approximately 40\% of the resolvers support UDP payload sizes of up to 512 bytes, while 50\% of the resolvers advertise a payload size equal or larger than 4000 bytes. The remaining 10\% are between 1232 and 2048 bytes. Given the minimum MTU size measurement of the nameservers in 1M-top Alexa domains in Figure~\ref{fig:alexa_top_1m_domains_fragsize}, this means that the resolver population is essentially portioned in two groups: one group (40\%) which is vulnerable to poisoning attacks with 7\% of all vulnerable domains and one group (50-60\%) which is vulnerable to poisoning attack with all the vulnerable domains.

\subsection{Comparative Analysis}
\label{sc:dns:method-eval}
Our measurements show that the methodologies for DNS cache poisoning can often result in practical attacks, depending on the setup, network conditions and server configurations. In this section we compare the DNS cache poisoning methodologies with respect to stealthiness, effectiveness and applicability. %

The main insights of the experimental measurements that we performed using each of the methods in Section \ref{sc:dns:poison} are summarised in Table~\ref{tab:method_comparison}. The columns in Table \ref{tab:method_comparison} correspond to the attacks we carried out against the domains and resolvers in our dataset (see Section~\ref{sc:eval}). 
\begin{table}[ht!]
    \centering
    \small

    \setlength\tabcolsep{2.0pt}
    \begin{tabular}{|r|c|c|c|c|c|}
    \hline
                    & \mcc{2}{\textbf{BGP Hijack}}   &\mr{1}{\textbf{SadDNS}}& \mcc{2}{\textbf{Fragmentation}}  \\ \cline{5-6} \cline{2-3}
    
                    & sub-      & same-     &              & any IPID     & global IPID \\ \hline \hline
    \multicolumn{6}{|c|}{ \textbf{Applicability} } \\ \hline
    \makecell[r]{Vuln. resolvers\vspace{2pt}} & 70\% \hspace{-17pt} \mr{2}{\scriptsize or}        & 80\% \hspace{-17pt} \mr{2}{\scriptsize or}  & 11\% \hspace{-18pt} \mr{2}{\scriptsize and}         &\mcc{2}{ \hspace{15pt} 91\%             \hspace{-40pt} \mr{2}{\scriptsize and \hspace{30pt} and} }\\ %
    Vuln. domains   & 53\%     & 70\%       & 12\%         & \multicolumn{1}{c}{4\%} & \multicolumn{1}{c|}{1\%}      \\  \hline \hline
    
    \multicolumn{6}{|c|}{ \textbf{Effectiveness} } \\ \hline
    Hitrate         &\mcc{2}{ 100\%        }& 0.2\%        & 0.1\%    & 20\%   \\ \hline
    Queries needed  &\mcc{2}{ 1        }    & 497          & 1024     & 5   \\ \hline
    Total traffic (pkts)   &\mcc{2}{ 2     }& 987K         & 65K          & 325        \\ \hline \hline
    
    \multicolumn{6}{|c|}{ \textbf{Stealthiness} } \\ \hline
    Visibility      & \makecell{very\\visible} & visible & \mcc{2}{\makecell{stealthy, but locally de-\\tectable (Packet flood)}} & very stealthy \\ \hline \hline

    \multicolumn{6}{|c|}{ \textbf{Additional requirements} } \\ \hline
    Additional     &\mcc{2}{\mr{2}{ none }}& \mr{2}{none} &\mcc{2}{ max(resolver EDNS size) }\\
    requirements   &\mcc{2}{}              &              &\mcc{2}{ $<$ min(nameserver MTU) }\\

    \hline
    
    \end{tabular}

    \caption{Comparison of the cache poisoning methods.}
    \vspace{-20pt}
    \label{tab:method_comparison}
\end{table}

\begin{figure}
    \centering
    \begin{subfigure}[b]{0.23\textwidth}
        \centering
        \includegraphics[width=\textwidth]{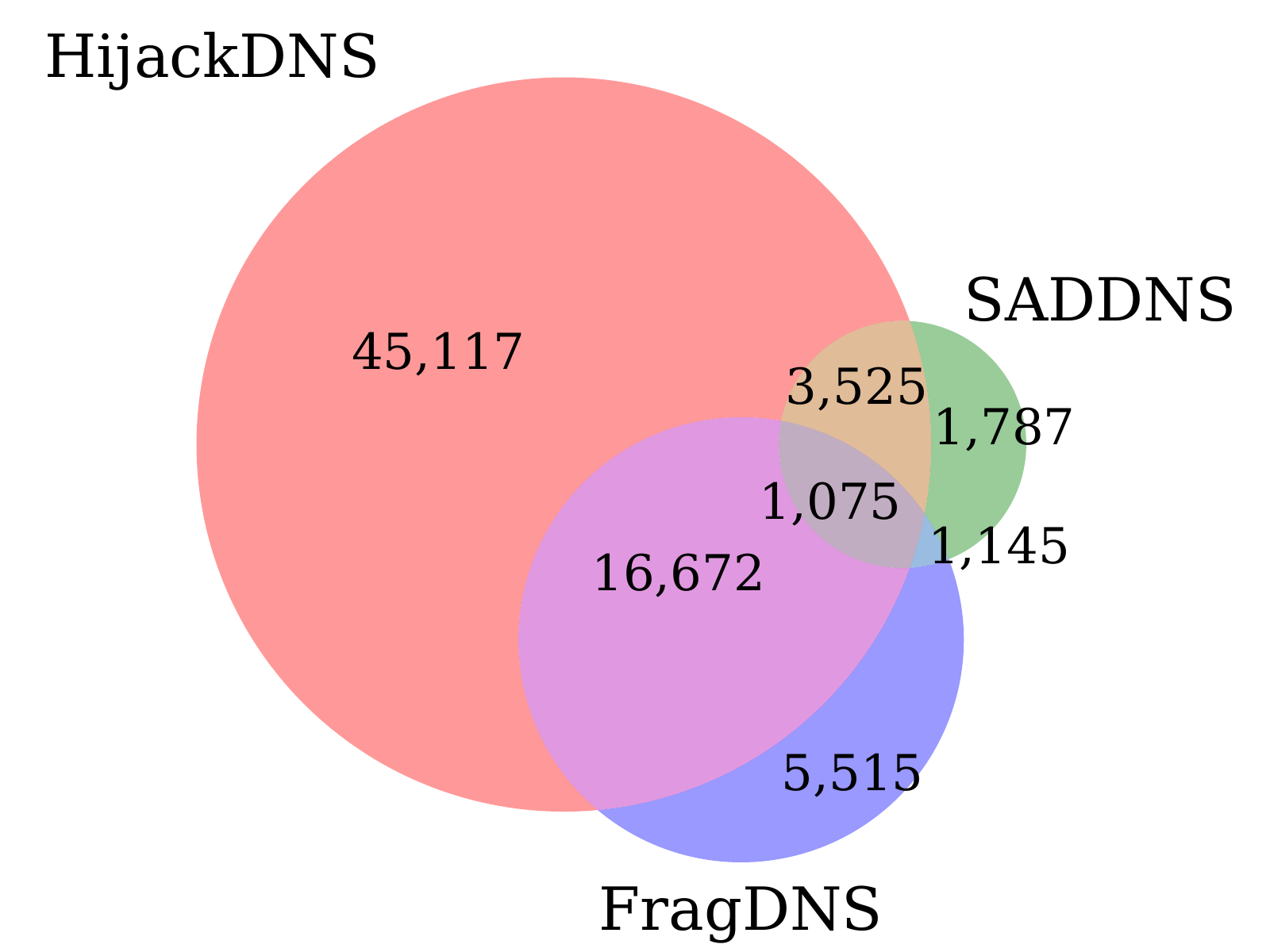}
        \caption{Resolver}
        \label{fig:vuln_resolvers_venn}
    \end{subfigure}
    \hfill
    \begin{subfigure}[b]{0.23\textwidth}
        \centering
        \includegraphics[width=\textwidth]{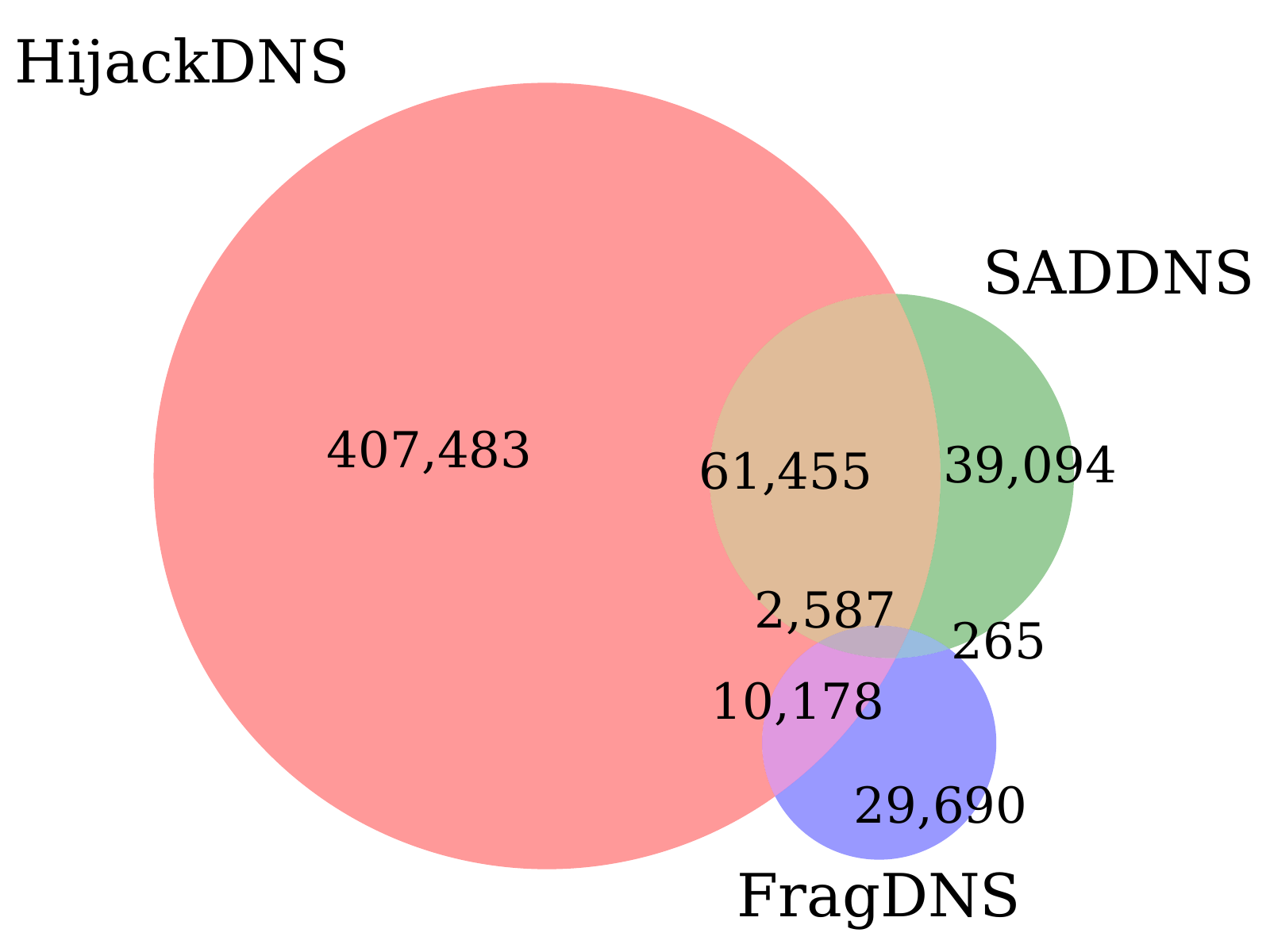}
        \caption{Domain}
        \label{fig:vuln_domains_venn}
    \end{subfigure}
    \caption{\begin{added}
    Venn diagram of all vulnerable resolvers (by number of back-end addresses) and domains.\end{added}}
    \label{fig:venn}
\end{figure}

\subsubsection{Applicability}%

\begin{added}
A method is applicable against a resolver for some domain if it results in a practical DNS cache poisoning attack. The applicability for each method for resolvers and domains is listed in Table \ref{tab:method_comparison}.

To compare the applicability of the methodologies we use the results of our internet measurements (Tables~\ref{tab:vulnerable_resolvers_summary} and \ref{tab:vulnerable_domains_summary}) and take the numbers for the ad-Net resolvers and 1M-top Alexa domains datasets. We also show the absolute number of all vulnerable resolvers (according to a back-end address) and domains in all our datasets in Figure~\ref{fig:venn}. This figure shows that the number of resolvers and domains vulnerable to \hijack{} is by far the highest, while \sad{} has more vulnerable domains and \frag{} has more vulnerable resolvers. The overlaps between the vulnerable domains and resolvers can be seen as expected for a distribution of unrelated properties, i.e., \sad{} and \frag{} have a significant overlap with \hijack{}, which is due to the fact that 53-70\% of the systems we measured are vulnerable to \hijack{}, while \sad{} and \frag{} only have a small overlap compared to number of vulnerable systems in each category. Only 11\% of the DNS resolvers and 12\% of the domains are vulnerable to \saddns\ attack. Many more resolvers are vulnerable to injection of content via IPv4 fragments, hence \frag\ attack is more applicable than \sad. In addition, due to its large size, the open resolver dataset dominates the results in our comparison.

\end{added}

\subsubsection{Effectiveness} Attack effectiveness is demonstrated with the traffic volume needed for a successful attack, which is a function of the number of queries that should be triggered for a successful attack. The larger the attack volume, the less stealthy the attack is. We define hitrate as the probability to poison the target DNS cache with a single query and calculate the expected number of queries for each of the poisoning methods by inversion of the hitrate. We estimate the expected number of packets sent to the resolver by multiplying this with the traffic volume generated per query. For \sad\ where the amount of traffic during the attack is not stable, we analyse the experimental data for the amount of traffic needed.

\hijack. If an AS prefers a malicious BGP announcement of the adversary to the announcement of the victim AS, then the attack is effective, requiring only a single packet to send a malicious BGP announcement and then another packet to send a spoofed DNS response with malicious DNS records. 

\sad. Using our implementation of \sad\ attack from Section \ref{sc:saddns:eval} we find that the DNS cache poisoning with \sad\ succeeds after an average of 471 seconds (min: 39 seconds, max: 779 seconds). This is inline with the results in \cite{saddns} which report an average of 504 seconds. To achieve a successful attack we needed to run 497 iterations on average. This is correlated with the attack duration since we do not trigger more than two queries per second. When more queries within one attack iteration are triggered, the resolvers respond with {\tt servfail}. %
 By inverting this number we get a hitrate of 0.2\%. Notably however, since most of the queries do not result in attack windows of meaningful length, an attacker should be able to optimise the attack by analysing the exact back-off strategies used by the target resolver, and adjusting the queries according to this. 
 
Using the results from our \sad\ experiment, we also obtain statistics for how may packets are sent to the target resolver. On average, our implementation sent 986,828 packets or 88MB of traffic, which is again, comparable to the original attack (69MB in \cite{saddns}).

\frag. Only about 1\% of the domains allow deterministic fragmentation-based cache poisoning attacks thanks to slowly incremental global IPID counter in nameservers. More than 4\% of the domains are vulnerable to probabilistic attacks by attempting to hit an unpredictable IPID counter and to match the UDP checksum. When the IPID values are not predictable, the probability to hit the correct value is roughly $0.1\%$.
To match the UDP checksum, the attacker needs to predict the partial UDP checksum of the second fragment of response sent by the nameserver. This means that the probability to match the UDP checksum is the inverse of the number of possible second fragments emitted by the nameserver (assuming equal distribution). 

To calculate the per-nameserver hitrate of \frag\ attack for each domain we calculate the product of both probabilities, matching the IPID as well as matching the UPD checksum. We take the average of these per-nameserver hitrates to calculate a per domain hitrate. 
The results of our evaluation are: when the nameservers use a single global counter for IPID, depending on the rate at which queries arrive at the nameserver, the median hitrate over all vulnerable domains (for different rates of queries from other sources) is 20\%. When the nameserver selects IPID values pseudorandomly, the median hitrate is 0.1\% which is the probability to correctly guess the IPID, as most servers to not randomise the records in DNS responses.

\frag\ attack also requires large traffic volumes with 1024 packets median computed over vulnerable domains with 65K packets for an unpredictable IPID, and with only 325 packets on average against a predictable IPID against high load servers, such as the servers of top-level domains.

In the worst case, the attack requires 64 packets to fill the resolver IP-defragmentation buffer and another packet to trigger the query. Combined with a 0.1\% success rate, this translates to an average of 65,000 packets.

\subsubsection{Stealthiness} In BGP prefix hijacks malicious BGP announcements manipulate the control plane and a single BGP announcement suffices to change the forwarding information in the routers. BGP prefix hijacks generate lower traffic volume when performing the hijack but may be more visible in the Internet since the attack impact is more global. The more networks are affected as a result of the BGP hijack the higher the chance is that such attacks may be detected. Same-prefix hijack is more stealthy in control plane than sub-prefix hijack since it does not affect the global routing BGP table in the Internet, but causes manipulations only locally at the ASes that accept the malicious announcement. Furthermore, as we already mentioned, short-lived BGP hijacks typically are ignored and do not trigger alerts \cite{boothe2006short,karlin2008autonomous,khare2012concurrent}. %
In contrast, guessing the source port with \sad\ method (Section \ref{sc:saddns}) or injecting malicious payload via IPv4 fragmentation (Section \ref{ssc:frag}) generate more traffic than BGP hijacks, but only locally on the network of the victim DNS resolver or the target nameserver. In contrast to BGP hijacks the attack is performed on the data plane, and is hence not visible in the global BGP routing table in the Internet. 

 \saddns\ attack creates a large traffic volume and hence may be detected by the affected networks. \frag\ attacks against domains that uses a global sequentially incremental IPID counter are the stealthiest.

\section{Countermeasures}\label{sc:defences}

Almost all Internet systems, applications and even security mechanisms use DNS. As we showed, a vulnerable DNS introduces not only threats to systems using it but also to security mechanisms, such as PKI. We provide recommendations to mitigate that threat. %

We also set up a tool at \url{https://crosslayerattacks.sit.fraunhofer.de} to allow clients to check if their networks are operating DNS platforms vulnerable to the cache poisoning attacks evaluated in our work.
In the rest of this section we separately explain our recommendations for DNS servers to prevent cache poisoning attacks and then for applications to prevent cross-layer attacks.

\subsection{DNS servers} In addition to recommendations and best practices for patching DNS servers, such as those in [RFC5452] \cite{rfc5452}, we recommend a new countermeasure we call {\bf security by obscurity}. Our experience of cache poisoning evaluation in the Internet showed that the less information the adversary has, the more hard it becomes to launch the attacks in practice. Security by obscurity proves effective not only against off-path but also against on-path MitM attacks. Although it is a known bad practice in cryptography it turns out useful in practice. Specifically, for launching the attacks the attackers need to collect intelligence about the target victims, such as which caching policies are used, which IP addresses are assigned to the resolver - randomising or blocking this information, will make a successful attack harder. The network administrators can deploy countermeasures to make such information difficult to leak, e.g., DNS resolvers should use multiple caches with different DNS software on each, resolvers should not send ICMP errors, nameservers should randomise records in responses.

\begin{added}

{\bf Preventing queries.} Server operators might choose to configure systems to do less (or no) DNS lookups, ie. in the case of email servers. This reduces the chance an attacker can trigger a query to start the poisoning.

{\bf Blocking fragmentation.} Resolver operators can block fragmented responses in firewalls to reduce the applicability of \frag{} attacks. Some operators only implement filtering of small fragments (i.e., Google's 8.8.8.8) which can prevent the attack since the attacker might not be able to cause a nameserver response of the size needed to reach the filtering limit.

{\bf Randomise DNS responses.} Randomising nameserver responses complicates the \frag{} attack as the attacker needs to predict the UDP checksum of the original nameserver's response.

{\bf 0x20 Encoding.} 0x20 Encoding adds entropy to the DNS query which must be matched by the response. This complicates the \sad{} attack to a point where it is no longer viable (ie. adding 0x20 Encoding to a domain with 16 alphanumeric characters adds 16 bits of entropy to the query). Since this randomness is only contained in the question section of the DNS packet, it cannot prevent the \frag{} attack as it will be in the first fragment along with the TXID.

{\bf Securing BGP.} Full deployment of RPKI (together with BGPSec) would prevent the \hijack{} attack. However, because of several deployment barriers, most of the prefixes are not protected by RPKI and most ASes do not enforce Route Origin Validation (ROV) \cite{gilad2017we,DBLP:conf/dsn/HlavacekHSW18,DBLP:journals/corr/0002BCKSW17}. We refer to \cite{hlavacek2020disco} for a comprehensive discussion of the deployment issues.

\end{added}

\subsection{Applications}
In the rest of this section we provide recommendations for preventing cross-layer attacks that use DNS cache poisoning. %

{\bf Separate resolvers and caches.} It is common in networks to use one DNS resolver for multiple services and servers. Our attacks exploit that. We recommend using different DNS resolvers (each with a distinct cache) for each system.

{\bf Third party authentication (TLS).} Third party authentication, like TLS, \begin{added}can mitigate the attacks against all DNS use-cases which aim to locate a server (i.e.m federation and address lookup use-cases).\end{added} However, even such mechanisms can only reduce the harm of DNS poisoning, but not completely mitigate it, e.g., adversaries can use DNS cache poisoning to subvert the security of DV during certificates issuance. \begin{added} Furthermore, an attacker can still use cache-poisoning for DoS attacks.\end{added}

{\bf Two factor authentication.} Should be enabled by default (and not optional as it is now). This would prevent the attacker from getting access to the account even if it has acquired the login credentials for the victim.

\begin{added}
{\bf Secure fallback.} Instead of allowing a transaction when no information about its authorisation state can be gathered (like done currently in SPF and RPKI) a security-mechanism could decide to not allow it. This however would mean that attacking the availability of DNS for a certain domain would allow DoS attacks instead, preventing a resolver from looking up a domain's SPF records would prevent that domain from sending any emails to the servers using this resolver.
\end{added}

\section{Conclusions}\label{sc:conc}
We evaluated methodologies for launching practical DNS cache poisoning attacks and derived insights on the applicability, effectiveness and stealth of these attacks. We then applied the methodologies for a systematic evaluation of cross-layer attacks against popular applications. 

Our work demonstrates the significant role that DNS plays in the Internet for ensuring security and stability of the applications and clients. If DNS is vulnerable, our work shows that in addition to traditional attacks, such as redirection to adversarial hosts, weak off-path adversaries can even downgrade protection of security mechanisms, such as RPKI or DV. We provide recommendations for mitigations and developed a public tool at \path{https://crosslayerattacks.sit.fraunhofer.de} to enable clients to identify vulnerabilities in DNS platforms on their networks.

\section*{Acknowledgements}
We thank the anonymous referees for thoughtful feedback on our work. This work has been co-funded by the German Federal Ministry of Education and Research and the Hessen State Ministry for Higher Education, Research and Arts within their joint support of the National Research Center for Applied Cybersecurity ATHENE and by the Deutsche Forschungsgemeinschaft (DFG, German Research Foundation) SFB~1119.

\bibliographystyle{ACM-Reference-Format}
\bibliography{refs,NetSec,newrfc}

\balance
\end{document}